\documentstyle[12pt,epsfig]{./ioplppt}
\begin{document}
\jl{2}

\title{Friedel oscillations in a gas of interacting one-dimensional
fermionic atoms confined in a harmonic trap}[Friedel oscillations in a harmonic trap]

\date{\today}

\author{S N Artemenko\dag, Gao Xianlong\ddag, and W Wonneberger\ddag}

\address{\dag Institute of Radio Engineering and Electronics, Moscow 125009, Russia}

\address{\ddag\ Abteilung Mathematische Physik, Universit\"at Ulm, D89069 Ulm, Germany}

\begin{abstract}
Using an asymptotic phase representation of the particle density operator $\hat{\rho}(z)$
in the
one-dimensional harmonic trap, the part $\delta \hat{\rho}_F(z)$ which describes the
Friedel oscillations is extracted. The expectation value $\langle\delta \hat{\rho}_F(z)
\rangle$ with respect to the interacting ground state requires
the calculation of the mean square average of a properly defined phase operator.
This calculation is performed analytically for the Tomonaga-Luttinger model with
harmonic confinement. It is found that the envelope of the Friedel oscillations at zero
temperature decays with the boundary exponent $\nu = (K+1)/2$ away from the classical
boundaries. This value differs from that known for open boundary conditions or strong
pinning impurities. The soft boundary in the present case thus modifies the decay of
Friedel oscillations. The case of two components is also discussed.
\end{abstract}


\maketitle

\section{Introduction}

Recent experimental successes in obtaining degeneracy in three-dimensional ultracold Fermi
vapors \cite{DMJ99,OHara00,SFCCKMS01,Tru01,SK01,H02,H03}, possibly in combination with 
microtrap
technology \cite{VFP98,FGZ98,DCS99,RHH99,Ott01,L02}, make it conceivable to realize the
quasi one-dimensional neutral Fermi gas confined in a trapping potential without the
complications due to contacts and impurities.

Friedel oscillations are a principal feature of a degenerate Fermi gas
when translational invariance is broken. Usually, impurities \cite{F58}
are the cause.
However, boundaries can also be responsible for Friedel oscillations.
The spatial period of Friedel oscillations is $\lambda _F= \pi/k_F$, where
$k_F$ is the Fermi wave number.
This effect is particularly pronounced in one spatial dimension
because then the susceptibility  becomes
logarithmically  singular at $2 k_F$ due to perfect nesting.

It is known from the theory of one-dimensional fermions confined between 
hard walls (bounded Luttinger liquids = BLL) that interactions
modify the decay of Friedel oscillation away from the boundary \cite {FG95,WVP96,VYG00}.

In this article, we investigate the Friedel oscillations at zero temperature 
both for the spin-polarized one component system, when s-wave scattering is forbidden, 
as well as for the two component system.
We apply a recent model of $N \gg 1$ interacting fermionic atoms in one spatial
dimension confined in a harmonic trap \cite{WW}. The model which can be termed
"Tomonaga-Luttinger model with harmonic confinement" is analytically solvable in a similar way
as the Luttinger model (cf. e.g. \cite{E79,Haldane,V95,Sch95}) using bosonization. The present
model is simpler than the Luttinger model in that it has only one (non-chiral) branch in
contrast to the two chiral branches of the latter which result from an artificial split of an
otherwise continuous band.

It is found that near the classical boundaries, Friedel oscillations
decay in a way which differs from the known result for BLL \cite{FG95,WVP96,VYG00}.
It also differs from the result for an infinitely strong pinning
impurity \cite{EG95} which acts as an invariant hard wall under scaling \cite{KF92}.

The calculations become possible because the fermion density in the harmonic trap can be
decomposed asymptotically (i.e., for $N \gg 1$) into a slowly varying part and a part
describing the Friedel oscillations. Both parts involve a specific phase operator for which
a free field theory is available.

Friedel oscillations in realistic Fermi gas are difficult to observe
experimentally at least for two reasons:
the integrated total mass in the one-dimensional Friedel-oscillations 
is of the order of one atom (cf. e.g.\cite{GWSZ00}) though repulsive interactions increase
their weight. Furthermore, temperature effects
blur the oscillations unless $k_B T < \hbar \omega _\ell$ where $\omega _\ell$
is the longitudinal trap frequency \cite{AVMT02}. Provided these exceedingly low 
temperatures can be realized, we can adopt the argument in \cite{GWSZ00}:
it is conceivable to use an array of short micro traps each filled with a
reduced number of atoms (thus avoiding instabilities).  The oscillations within each trap 
then add up
and lead to a total effect that is enhanced by the number of traps.
Using micro fabrication techniques it should be possible to combine 100
traps on one substrate leading to a signal that may become within reach of
advanced imaging techniques.

\section{Theoretical framework}

\subsection{Tomonaga-Luttinger model with harmonic confinement}

Here, we give a short review of the model which is used for the present calculation.
More details are given in \cite{WW}. 

We start from an effective pair interaction

\begin{eqnarray}\label{0}
\hat{V} =\frac{1}{2} \sum _{mnpq}
V(m,p;q,n)\,(\hat{c}^+_m \hat{c}_q ) ( \hat{c}^+_p \hat{c}_n )
\end{eqnarray}
between one-dimensional spin polarized fermions. The fermionic creation and destruction  operators
$\hat{c}^+_m$ and $\hat{c}_q$ are taken in the basis of harmonic oscillator wave functions.
Thus the harmonic trap topology is exactly represented.

The Hamiltonian for interacting  fermions considered here follows uniquely 
by retaining those parts in the fermionic pair interaction operator
which are expressible in terms of density fluctuation operators $\hat{\rho}(m)=\sum _p
\hat{c}^+_{p+m}\hat{c}_p$. These are

\begin{eqnarray}\label{00}
\fl V (m,p;q,n)  \rightarrow  V _a(|q-m|) \, \delta _{m-q, n-p}+V
_b(|q-m|) \, \delta _{q-m, n-p}
%\\[2mm]\nonumber
+ V _c(|q-p|) \, \delta _{m+q, n+p}.
\end{eqnarray}

The interaction matrix elements $V_a(m)$, $V_b(m)$, and $V_c(m)$ correspond to the
Luttinger model coupling functions $g_4(p)$, $g_2(p)$, and $g_1(p)$, respectively. $V_a$
and $V_b$ describe forward scattering and $V_c$ describes $2 k_F$ (backward) scattering.
In \cite{GW02}, it is shown that the retained matrix elements are dominant in the limit
of large $N$. This is related to approximate momentum conservation during collisions 
in the trap.

In the next step, the linear dispersion of free harmonic oscillator states and
the addition of the anomalous vacuum is utilized to bosonize the original fermionic
Hamiltonian in terms of canonically conjugate Bose operators $\hat{d}$ and
$\hat{d}^+$ in accordance with Kronig's identity \cite{deK35}. This gives the bosonic form 

\begin{eqnarray}\label{1}
 \tilde{H} & = & \frac{1}{2} \,\hbar \omega _\ell \, \sum _{m>0} m \left(
 \hat{d}_m \hat{d}_m ^+ +  \hat{d}_m ^+ \hat{d}_m \right) - \frac{1}{2}
 \sum _{m>0}V_c(m)\, m
 \left ( \hat{d}_m^2+ \hat{d}^{+2}_m \right)\\ \nonumber
&&+ \frac{1}{2} \sum _{m>0} V_c(m)\, \sqrt{2m} \left [ \hat{d}
_{2m} + \hat{d}^+ _{2m} \right ]
\\\nonumber
&&+ \frac{1}{2} \sum _{m>0} V_a(m)\, m \left(
 \hat{d}_m \hat{d}_m ^+ +  \hat{d}_m ^+ \hat{d}_m \right)
 +\frac{1}{2} \sum _{m>0} V_b(m)\, m
 \left ( \hat{d}_m^2+ \hat{d}^{+2}_m \right).
 \end{eqnarray}
of the Hamiltonian. 

Equation (\ref{1}) is the version for a one-component Fermi gas, i.e., for spin polarized
fermions. Usually, the remaining "p-wave" interaction is small. However, it has been 
demonstrated recently that Feshbach resonances can make it relevant \cite{R03}. 

Our simplified interaction Hamiltonian is integrable. In the sense of the Luttinger 
liquid phenomenology, we expect that the boundary exponent for Friedel oscillations
at zero temperature and for $N \gg 1$ is invariant against details of the interaction.
It is also stressed that the trap potential is exactly incorporated.

In the one-component system, backscattering dominates as demonstrated below. Accepting
this, the validity of our approach could be verified analytically by perturbation theory
in the fermionic Hilbert space as well as by exact numerical diagonalization of the 
fermionic problem \cite{WW}. 
%In the Appendix, we give another example of the validity by calculating the Friedel 
%oscillations directly in first order perturbation theory.

\subsection{Dominance of backward scattering in the one-component gas}

The existence of only one branch results in restrictions on the values of the
interaction coefficients $V_a$ and $V_b$ in the one-component case: their contribution is
small in comparison to $V_c$ \cite{GW02}. This can be demonstrated analytically by using WKB 
wave functions (cf. equation (\ref{8}) below) in the calculation of $V(m,p;q,n)$: 
starting from an effective "p-wave" potential in one dimension

\begin{eqnarray}\label{01}
V(z)=V_p \,a_p^3 \,\partial ^2_z \,\delta(z),
\end{eqnarray}
one obtains for $(m,n,p,q)=O(N) \gg 1$

\begin{eqnarray}\label{02}
V(m,p;q,n)= \int dz\,dz'\,\psi _m(z)\psi _q(z)\,V (z-z')\,
\psi _p(z') \psi _n(z')
\\[2mm]\nonumber 
=V_p \,a_p^3 \,\int dz\,\psi _m(z)\psi _q(z)\,\partial ^2_z\,\{\psi _p(z) \psi _n(z)\}
%\\ [2mm]\nonumber
\rightarrow \frac{4 \sqrt{2 N} \alpha^3 a_p^3 V_p}{\pi^2}\,F(s),
\end{eqnarray}
with

\begin{eqnarray}\label{03}
F(s) \equiv \frac{\cos^2(\pi s/2)}{s^2-1},\quad s \equiv m+q-p-n. 
\end{eqnarray}
Thus, each individual backscattering term in (\ref{00}) belongs to the dominating $s=0$ 
contribution while only some terms of $V_{a,b}$-type do this. Thus backscattering
dominates. 

The interaction coefficient $V_c =V_c(1)$ can be expressed as

\begin{eqnarray}\label{04}
V_c=-\frac{2}{\pi^2}\, k_F a_p \left(\frac{V_p}{\frac{\hbar^2}{2 m_A a_p^2}}\right)\,
\hbar\,\omega _\ell.
\end{eqnarray}

\subsection{One-particle operator and phase fields}

The third contribution on the r.h.s. of equation (\ref{1}) represents a one-particle operator
$\hat{V}_1$. It originates from rearranging operators in equation (\ref{0}) to bring 
backscattering into a bilinear form of density fluctuation operators.
The one-particle operator is neglected in the BLL as pointed out in \cite{MS00}. It
does not alter boundary exponents, but has quantitative effects on other properties.
$\hat{V}_1$ is exactly taken into account in the present model. 

The central dimensionless coupling constants and  the renormalized level spacings
for the model, equation (\ref{1}), are given by \cite{WW}

\begin{eqnarray}\label{2}
K_m & \equiv &\sqrt{\frac{\hbar \omega _\ell +
V_a(m) - (V_b(m)-V_c(m))}{\hbar \omega _ \ell + V_a(m) + (V_b(m)-
V_c(m))}}, \quad
\\ [2mm] \nonumber
\epsilon _m &\equiv& \sqrt{(\hbar \omega _\ell+V_a(m) )^2 -
(V_b(m)-V_c(m))^2},
\end{eqnarray}
respectively.

For simplicity, the dependence of $K$ and $\epsilon$ on $m$ is occasionally suppressed.
Note that $V_a(m) \rightarrow V_a$ implies that there are no interaction effects due to
this matrix element in the one component theory and $V_a$ is strictly zero while
$V_b$ can be neglected, i.e., the central coupling constant is

\begin{eqnarray}\label{3a}  
K =  \sqrt{\frac{\hbar \omega _\ell     
+V_c}{\hbar \omega _ \ell - V_c}},\quad \epsilon = \sqrt{(\hbar \omega _\ell )^2
 - V_c^2},
\end{eqnarray}
with $V_c$ given by equation (\ref{04}).

In some physical quantities, especially in the mean square phase fluctuation calculated
below, the neglect of the m-dependence leads to inconsistencies because
stability requires the coupling constants $K_m$ to approach unity for large m \cite{Haldane}.
In analogy to the Luttinger model, we write approximately

\begin{eqnarray}\label{3c}
K_m=1+(K-1)\,\exp(-m r).
\end{eqnarray}

An estimate for $r$ is $R/L_F \ll 1$ where $R$ is the spatial range of
the interaction.

Assuming the same exponential decay for the interaction matrix elements

 \begin{eqnarray}\label{3}
 V_c(m)= V_c \,\,e^{-m r},\quad r \ll 1,
 \end{eqnarray}
the one-particle operator can be rewritten as

\begin{eqnarray}\label{4}
\hat{V}_1=\frac{1}{4 \pi}\,V_ c \,\int _ {-\pi}^\pi du\,\left[\frac{e^{-r +2 i u}}
{1-e^{-r+2 i u}}
+\frac{e^{-r -2iu}}{1-e^{-r -2 i u}}\right]\,\partial _u \hat{\phi}_{odd}(u),
\end{eqnarray}
with the phase field

\begin{eqnarray}\label{4a}
\fl \hat{\phi}_{odd}(u) \equiv \frac{1}{2}\,(\hat{\phi}(u)+\hat{\phi}^+(u)-\hat{\phi}(-u)-
\hat{\phi}^+(-u))
= \sum _{n=1}^\infty \sqrt{\frac{e^{-n \eta}}{n}}\, 
\,\sin(n u)\,\left ( \hat{d}_n +\hat{d}^+ _n \right).
\end{eqnarray}
and $\hat{\phi}$ is the bosonization phase operator \cite{SchM96},

\begin{eqnarray}\label{4b}
\hat{\phi} (u) \equiv - i \sum ^\infty _{n=1} \frac{1}{\sqrt{n}}\, e ^{i n (u+i \eta/2)}
\,\hat{d}_n.
 \end{eqnarray}

In diagonalizing equation (\ref{1}), the one-particle operator contributes a c-number 
shift which is taken care of by the new phase operator

\begin{eqnarray}\label{4c}
\hat{\Phi}(u) = \hat{\phi}_{odd}(u)+b(u),
\end{eqnarray}
with

\begin{eqnarray}\label{5}
b(u)=i \,\frac{K V_c}{4 \epsilon }\ln\left(\frac{1-e^{-r+2 i u}}
{1-e^{-r -2 i u}} \right).
\end{eqnarray}

In contrast to $\hat{\phi}_{odd}$, the new phase field $\hat{\Phi}$
is homogeneous in the operators $\hat{f}$ and $\hat{f}^+$ which diagonalize the
Hamiltonian: $\hat{H}= \sum _m m \epsilon _m 
\hat{f}_m^+\hat{f}_m$ + zero mode contributions. The zero mode plays no role in the present 
context. The representation of the phase field $\hat{\Phi}$ in terms of the diagonalizing
operators is

\begin{eqnarray}\label{5a}
\hat{\Phi}(u)=\sum _{m=1}^\infty \sqrt{\frac{K_m}{m}}\,\,e^{-m \eta/2}\,\sin mu\,(\hat{f}_m
+\hat{f}^+_m).
\end{eqnarray}

\section{Decomposition of the density operator for the harmonic trap}

The density operator $ \hat{\rho}(z)$ in Fock space for fermions in the harmonic
trap is

\begin{eqnarray}\label{6}
\hat{\rho}(z)= \sum _{m=0,n=0}^\infty\psi _m(z)\psi _n(z)\,\hat{c}^+_ m
\,\hat{c}_n.
\end{eqnarray}

Bosonizing $\hat{\rho}(z)$ using the auxiliary field method in \cite{SchM96}, i.e., 

\begin{eqnarray}\label{7a}
\hat{c}^+ _m  \hat{c}_n  &=& 
\int^{2 \pi}_0 \int^{2 \pi}_0 \,\frac{du dv}{4 \pi^2} \,e^{i(mu-nv)}
\,\frac{e^{-i(N-1)(u-v)}} {1-e^{-\eta+i(u-v)}}\,
\\[4mm]\nonumber
&&\times\exp \left\{ - i \left( \hat{\phi}^+(u) - \hat{\phi}^+(v) \right) \right\}
\exp \left\{ - i \left( \hat{\phi} (u) -\hat{\phi} (v) \right) \right\}     
\end{eqnarray}
implies a projection onto the  subspace of $N$ fermions: $ \hat{\rho}(z)\rightarrow
\hat{\rho}_N(z)$.

We will extract from $\hat{\rho}_N(z)$ a slowly varying part and a part associated with
the Friedel oscillations.

After bosonization, $\hat{\rho}_N(z)$ is given by

\begin{eqnarray}\label{7}
\hat{\rho}_N(z) &=& \int _{-\pi}^\pi\,\frac{du dv}{4
\pi^2}\,\frac{e^{-i(N-1)(u-v)}} {1-e^{-\eta+i(u-v)}}\,
 e^{-i \hat{\phi}^+ (u)+i \hat{\phi}^+(v)}\,e^{-i \hat{\phi}(u)+i \hat{\phi}(v)}\,
\\ [2mm] \nonumber
&&\hspace*{3cm}\times\left[\sum _{m=0,n=0}^\infty\psi _m(z)\psi _n(z)\,e^{i m u-i n
v}\right].
\end{eqnarray}
Here, $\eta$ is a positive infinitesimal.

Using the form factor $Z(z)=\sqrt{1-z^2/L_F^2}$ of the harmonic trap, where
$2 L_F=2 \ell\sqrt{2 N-1}$ is the semi-classical extension of the Fermi sea,
the harmonic oscillator wave functions $\psi _m$ can be approximated for $N \gg 1$
by their WKB form ($\alpha=1/\ell$, $\ell$: oscillator length) according to

\begin{eqnarray}\label{8}
\psi _m(z) \rightarrow \left(\frac{2 \alpha^2}{\pi^2 m Z^2(z)}\right)^{1/4}\,\cos\left
(\int _0^z dx\,k_m(x)-\frac{\pi m}{2}\right).
\end{eqnarray}

Because of the rapidly oscillating phase factor
$\exp(-i(N-1)(u-v))$, an expansion around the Fermi edge with

\begin{eqnarray}\label{9}
\fl k_m(z)=\alpha\sqrt{2m+1-\alpha^2 z^2}\approx k_F Z(z)
+\frac{\tilde{m}}{L_F Z(z)},\,
%\\ [2mm] \nonumber
k_F=\alpha \sqrt{2 N-1},\, m=N-1 +\tilde{m}
\end{eqnarray}
and

\begin{eqnarray}\label{10}
\int _0^z dx\,k_m(x) & \approx &  \frac{k_F}{2}\,\left\{z Z(z)
+L_F\,\arcsin\left(
\frac{z}{L_F}\right)\right\}+\tilde{m}\,\arcsin\left(\frac{z}{L_F}\right)
\\ [2mm] \nonumber
& \equiv & k_F
\tilde{z}(z)+\tilde{m}\,\arcsin\left(\frac{z}{L_F}\right)
\end{eqnarray}
is reasonable because it will compensate that factor.
Explicitly, $\tilde{z}(z)$ is given by

\begin{eqnarray}\label{11}
\fl 2 k_F \tilde{z}(z) = k_F Z(z)z + k_F L_F \arcsin \frac{z}{L_F}
%\\ [2mm] \nonumber
 =  k_F Z(z) z + (2N-1)\arcsin \frac{z}{L_F}.
\end{eqnarray}

We consider the sum over $m$ in equation (\ref{7}). Extending the $\tilde{m}$-summation
to $-\infty$ and setting $m=N$ in phase insensitive terms, leads to the asymptotic
expansion

\begin{eqnarray}\label{12}
\sum _{m=0}^\infty \psi _m(z)\, e^{i m u}  \rightarrow 
\left(\frac{2 \pi^2 \alpha^2}{N Z^2(z)} \right)^{1/4} \,e^{i(N-1)u}
 \\[2mm] \nonumber 
\times \left\{ e^{i k_F \tilde{z}(z)-i \pi(N-1)/2}\left(1-i
\frac{\partial _u}{N}\right)^{-1/4}\, \delta(u+u_0(z))\right.
\\[2mm] \nonumber 
\left. +e^{-i k_F
\tilde{z}(z)+i \pi(N-1)/2}\left(1-i \frac{\partial
_u}{N}\right)^{-1/4}\,
 \delta(u-u_0(z))\right\}.
 \end{eqnarray}

It is noted that the core states are not properly represented in the expansion (\ref{12}).
As a consequence, one does not obtain correct information about the operator for the
average density. Thus we will retain only the fluctuating part $\hat{\rho}_N \rightarrow
\delta \hat{\rho}$ which is of main interest here.

Applying this expansion to equation (\ref{7}), we find to lowest order in $\partial _u/N$

\begin{eqnarray}\label{13}
\delta \hat{\rho}(z)= \frac{1}{\pi}\,\partial _z \hat{\phi}_{odd}(u_0(z))
 -\frac{(-1)^N}{\pi \eta L_F Z(z)}\,\cos 2[k_F \tilde{z}(z)+ \hat{\phi}_{odd}(u_0(z))].
 \end{eqnarray}

The phase field $\hat{\phi}_{odd}(u)$ is found to be identical to that of equation
(\ref{4a}). This phase operator $\hat{\phi}_{odd}$ plays a central role in the present 
investigation.

The density operator (\ref{13}) consists of two parts involving the phase operator
$\hat{\phi}_{odd}$, a gradient term for slow spatial
density variations and a rapidly oscillating term $\delta \hat{\rho}_F$ describing the Friedel
oscillations. This structure is well known from the theory of the Luttinger model. In the 
latter
case, the argument of the phase operator is the spatial coordinate. In the present case, 
however,
a non-linear relation between the spatial coordinate $z$ and the variable $u_0(z)$ in
the phase operator according to

\begin{eqnarray}\label{16}
u_0(z)= \arcsin\left(\frac{z}{L_F}\right)-\frac{\pi}{2},
\end{eqnarray}
appears which reflects the harmonic trap topology. Furthermore, it is a priori not clear what
the right structure of the phase operator in the confined case is. Our calculation gives
the answer in form of equation (\ref{4a}).

\section{Calculation of Friedel oscillations and boundary exponent}

In order to calculate the Friedel oscillations $\langle \delta
\hat{\rho}_F(z)\rangle _0$ in the interacting ground state, we
apply the Wick theorem in the well known form

\begin{eqnarray}\label{21}
\langle e^{i \hat{\Phi}}\rangle = \exp\left(-\frac{1}{2}\langle \hat{\Phi}^2 \rangle \right).
\end{eqnarray}
utilizing the homogeneous structure of the phase operator $\hat{\Phi}$ equation
(\ref{5a}). Thus

\begin{eqnarray}\label{22}
\fl\langle\cos 2[k_F \tilde{z}(z)-b(u_0(z)) +
\hat{\Phi}(u_0(z))]\rangle _0
%\\ [2mm] \nonumber
= \cos 2[k_F \tilde{z}(z)-b(u_0(z))]\,e^{-2 \langle
\hat{\Phi}(u_0(z))^2 \rangle _0 }.
\end{eqnarray}

The mean square average with respect to the ground state becomes

\begin{eqnarray}\label{24}
 \langle \hat{\Phi}(u)^2 \rangle _0=\sum _{m=1}^\infty \,\frac{\sin^2(m u)}{m}\,
\left(e^{- m \eta}+(K-1)\,e^{- m r}\right),
\end{eqnarray}
which leads to

\begin{eqnarray}\label{25}
\fl\langle \hat{\Phi}(u)^2 \rangle _0 = -\frac{1}{2}\,\ln \eta
&-&\frac{1}{2}\,(K-1)\,\ln r -\frac{1}{4}\,\ln[{\cal{D}}(2 u+i \eta)
{\cal{D}}(-2 u+i \eta)]
\\ [2mm] \nonumber
&& -\frac{1}{4}\,(K-1)\,\ln[{\cal{D}}(2 u+ir) {\cal{D}}(-2 u+i
r)].
\end{eqnarray}

Here, the abbreviation

\begin{eqnarray}\label{26}
{\cal{D}}(s) \equiv \frac{1}{1-e^{is}}
\end{eqnarray}
is introduced. Finally, one obtains

\begin{eqnarray}\label{28}
\fl\exp\{-2 \langle \hat{\Phi}(u)^2 \rangle _0 \}=\eta \,r^{K-1}\,2^{-K/2}\,
\frac{[(1+e^{-2 r})/2-e^{-r}\,\cos(2 u)]^{(1-K)/2}}
{[(1+e^{-2 \eta})/2-e^{-\eta}\,\cos(2 u)]^{1/2}}.
\end{eqnarray}

Considering $r \ll 1$, $\eta \rightarrow 0+$, and

\begin{eqnarray}\label{29}
\cos 2 u_0(z)= 2 \left(\frac{z}{L_F}\right)^2 -1,
\end{eqnarray}
leads to the result for the Friedel oscillations in the limit $L_F-|z|\gg r^2 L_F/8$

\begin{eqnarray}\label{30}
\langle \delta \hat{\rho}_F(z)\rangle _0 = -\frac{(-1)^N}{2 \pi L_F }\,
\left(\frac{r}{2}\right)^{K-1}\,\frac{
\cos 2[k_F \tilde{z}(z)-b(u_0(z))]}{Z(z)^{K+1}}.
\end{eqnarray}

It is seen that attractive interactions, $K >1$, decrease the Friedel oscillations,
while repulsive interactions, $K <1$, increase them at any fixed position $|z| < L_F$. 

The backscattering phase shift in equation (\ref{30}) is mostly a small correction as 
is seen by comparing equations (\ref{5}) and (\ref{11}), except when the coupling is 
exceedingly strong.

It is noted that the coupling constant $K$ goes to unity and $b(u_0(z))$ to zero for 
vanishing interactions. This answers a question put forward in \cite{GWSZ00}: the divergence 
of the envelope near the boundaries for free Friedel oscillations in the harmonic trap
has the conjectured boundary exponent $K_0=1$.
For $|z| \ll L_F$, i.e., well inside the trap, one obtains the free Friedel 
oscillations

\begin{eqnarray}\label{34}
\langle \delta \hat{\rho}_F(z)\rangle _{00} = -\frac{(-1)^N}{2 \pi L_F}\,\cos 2 k_F z,
\end{eqnarray}
in accordance with a corresponding result in \cite{GWSZ00}, which
was obtained by an asymptotic expansion of the exact particle density of non-interacting
fermions. 

Returning to the interacting case and specializing to the region near the classical 
boundaries, where

\begin{eqnarray}\label{31}
Z^2(z) \rightarrow 2\,(1-|z|/L_F),
\end{eqnarray}
the boundary exponent for the decay of Friedel oscillations away from
these classical turning points is

\begin{equation}\label{32}
\nu=(K+1)/2.
\end{equation}

In contrast, the corresponding
result for BLL in the case of one component is $\nu _{BLL}=K$ \cite{FG95,WVP96,VYG00}. 
The latter value is also obtained
for an infinitely strong pinning impurity which acts as a hard wall \cite{EG95}, a case 
when pinning strength does not scale \cite{KF92}. The present
soft boundary thus causes a slower decay for repulsive interactions in
comparison to the BLL while the situation is reversed for attractive interactions.

\subsection{Dependence on trap parameters}

We discuss the question how trap parameters influence the value of the
boundary exponent and the amplitude of Friedel oscillations. In view of $k_F=
\sqrt{m_A \omega _\ell (2N-1)/\hbar}$ and equation (\ref{04}), the central coupling
constant $K$ according to equation (\ref{3a}) depends on longitudinal trap frequency 
$\omega _\ell$ and particle number $N$ (apart from interaction data). 
Two cases will be considered:

\begin{itemize}
\item[i] The trap is made shallower, i.e., $\omega _\ell$ is decreased keeping N 
constant. In this case, interactions become irrelevant and $K$ goes to unity.
Because the amplitude in equation  (\ref{30}) is proportional to $1/L_F \propto
\sqrt{\omega _\ell}$ the Friedel oscillations vanish everywhere inside the huge
trap as expected. However, they still increase towards the boundaries because the 
trap topology persists for all non-zero $\omega _\ell$.

\item[ii]
A kind of thermodynamic limit \cite{Dalfovo} is defined by making 
$\omega _\ell \propto 1/N$ and $N$ large such that
the Fermi wave number and hence $\tilde{V}_c$ and $K \neq 1$ remain constant. Again,
the prefactor in equation (\ref{30}) suppresses the Friedel oscillations everywhere
inside the trap. Interestingly, at large but finite $N$ the amplitude of Friedel 
oscillations increases faster towards the boundary when interactions are attractive
in comparison to the repulsive case. 

\end{itemize}

\section{Boundary exponents for two components}

We develop the corresponding theory for two components, e.g., two different
hyperfine components of the same trapped fermionic atoms. We assume equal masses 
and equal trapping frequencies. The latter assumption is an approximation in the 
hyperfine case.

The local densities for mass and composition:

\begin{eqnarray}\label{35}
\hat{\rho}(z) \equiv \sum _{m=0,n=0}^\infty \psi _m(z) \psi
_n(z)\,[\hat{c}^+_{m+}\, \hat{c}_{n+}+\hat{c}^+_{m-}\,\hat{c}_{n-}
],\quad
\\ [2mm] \nonumber
\hat{\sigma}(z) \equiv \sum _{m=0,n=0}^\infty \psi _m(z) \psi
_n(z)\,[\hat{c}^+_{m+} \,\hat{c}_{n+}-\hat{c}^+_{m-}\,\hat{c}_{n-}
]
\end{eqnarray}
are treated as described in Section III. Two odd phase operators
corresponding to $\hat{\phi}_{odd}$ appear. They are defined by

\begin{eqnarray}\label{36}
\hat{\phi}_{\sigma,odd}(u)=\frac{1}{2}\,\left(\hat{\phi}_\sigma(u) +\hat{\phi}^+_\sigma(u)
-\hat{\phi}_\sigma(-u)-\hat{\phi}^+_\sigma(-u)\right),
\end{eqnarray}
in terms of the two bosonization operators for the components $\sigma = 1$ and $\sigma =-1$

\begin{eqnarray}\label{37}
\hat{\phi} _\sigma (u) = - i \sum _{m=1} \frac{1}{\sqrt{m}}\, e^{ im (u+i \eta/2)}\,
\hat{b}_{m \sigma}.
\end{eqnarray}

The relation of the $\hat{b}$-operators to mass and composition fluctuation operators $\hat{d}$
is

\begin{eqnarray}\label{38}
\hat{b}_{m \sigma} = \sum _\nu
\frac{1}{\sqrt{2}}\,\sigma ^{\frac{1-\nu}{2}}\,\hat{d} _{m \nu}
%\\[4mm]\nonumber
= \frac{1}{\sqrt{2}} ( \hat{d}_{m +} + \sigma \hat{d}_{m -}).
\end{eqnarray}

The final representation of the total density operator is

\begin{eqnarray}\label{39}
\fl\delta \hat{\rho}(z)= \frac{1}{\pi}\,\partial _z \left( \hat{\phi}_{+,odd}(u_0(z))
+\hat{\phi}_{-,odd}(u_0(z))\right)
\\ [2mm] \nonumber
-\frac{(-1)^N}{\pi \eta L_F Z(z)}\,\Bigg\{\cos \left [2 k_F \tilde{z}(z) +2 \hat{\phi}_{+,odd}(u_0(z))\right]
+ \cos\left[2 k_F \tilde{z}(z) +2 \hat{\phi}_{-,odd}(u_0(z)) \right]\Bigg\}.
\end{eqnarray}

Defining the basic phase fields

\begin{eqnarray}\label{40}
\hat{\Phi}_\nu(u) \equiv \frac{1}{\sqrt{2}}\,\{\hat{\phi}_{+,odd}(u)+\nu \,
  \hat{\phi}_{-,odd}(u)\}+b_\nu(u)
\end{eqnarray}
with $\nu=1$ for mass and $\nu=-1$ for composition fluctuations, brings equation (\ref{39}) 
into the form

\begin{eqnarray}\label{41}
\fl\delta \hat{\rho}(z)= \frac{\sqrt{2}}{\pi}\,\partial _z \, (\hat{\Phi}_{1}(u_0(z))
-b_1(u_0(z)))
\\[2mm] \nonumber
-\frac{2(-1)^N}{\pi \eta L_F Z(z)}\,\cos \left [2 k_F \tilde{z}(z)
+ \sqrt{2}\,(\hat{\Phi}_{1}(u_0(z))-b_1(u_0(z)))\right]
\cos(\sqrt{2}\,\hat{\Phi}_{-1}(u_0(z))).
\end{eqnarray}

The analogue of equation (\ref{5}) is

\begin{eqnarray}\label{42}
b_1 (u) = i \frac{K_1 V_{c \parallel}}{4 \epsilon _1}\, \sqrt{2}\ln \left
(\frac{1-e^{-r + 2 i u}}{1-e^{-r - 2 i u}} \right).
\end{eqnarray}

The quantity $b_{-1}$ vanishes identically.

Coupling constants and renormalized level spacings are given by

\begin{eqnarray}\label{43}
\fl K _{m,\nu} = \sqrt{\frac{(\hbar \omega _\ell + V_{a \parallel}(m)+ \nu V_{a \perp}(m))
-[ V_{b \parallel}(m)+ \nu V_{b \perp}(m)-V_{c \parallel}(m)]}{(\hbar \omega _ \ell +
V_{a \parallel}(m)+ \nu V_{a \perp}(m))+ [V_{b \parallel}(m)+\nu V_{b \perp}(m)-
V_{c \parallel}(m)]}},
\end{eqnarray}
and

\begin{eqnarray}\label{44}
\fl \epsilon _{m,\nu} = \left[(\hbar \omega _\ell + V_{a \parallel}(m)+ \nu V_{a \perp}(m))^2
 -[V_{b \parallel}(m) + \nu V_{b \perp}(m)- V_{c \parallel}(m)]^2 \right]^{1/2},
\end{eqnarray}
respectively. The subscript $\parallel$ refers to interactions between fermions of the
same component while $\perp$ stands for different components \cite{WW}.

Similarly, the composition operator reads as

\begin{eqnarray}\label{45}
\fl\delta \hat{\sigma}(z) = \frac{\sqrt{2}}{\pi}\,\partial _z \,
\hat{\Phi}_{-1}(u_0(z))
\\ [2mm] \nonumber
 +\frac{2(-1)^N}{\pi \eta L_F (z)}\,\sin \left [2 k_F
\tilde{z}(z) +
\sqrt{2}\,(\hat{\Phi}_{1}\left(u_0(z))-b_1(u_0(z))\right)\right]\,
 \sin(\sqrt{2}\,\hat{\Phi}_{-1}(u_0(z))).
\end{eqnarray}

The central relation equation (\ref{4c}) generalizes to

\begin{eqnarray}\label{46}
\fl \hat{\Phi}_\nu(u)  \equiv  \sum _{n=1}^\infty \sqrt{\frac{1}{n}}\,e^{-n \eta/2}\, \sin(n u)\,
\left ( \hat{d}_{n \nu} +\hat{d}^+ _{n \nu} \right)+b_\nu(u)
\\ [2mm] \nonumber
= \sum _{n=1}^\infty \sqrt{\frac{K_{n \nu}}{n}}\,\,e^{-n
\eta/2}\,\sin nu\,(\hat{f}_{n \nu} +\hat{f}^+_{n \nu}).
\end{eqnarray}

In order to be able to evaluate the phase fluctuations in the way given above, i.e., for free
phase fields, the interaction coefficient $V_{c \perp}$ for backward scattering between the
two components must be zero. This is usually not
the case. Relying on insight from the Luttinger model \cite{V95}, one can expect that for
$V_{c \parallel} \ge |V_{c \perp}|$ the coupling $V_{c \perp}$ scales to zero at low energies
and that for $V_{c \parallel} = V_{c \perp} > 0$ backscattering becomes irrelevant. In that
case, $K^*_{-1}=1$ holds. This corresponds to spin isotropy in the Luttinger model. However,
the present one-branch model hardly allows such a case in reality
due to suppression of s-wave scattering in the parallel channel. With this provision, we give
the results for two components based on the assumption that the field $ \hat{\Phi}_{-1}(u)$
is a (renormalized) free field:

\begin{eqnarray}\label{47}
\fl \langle \delta  \hat{\rho}_F(z) \rangle _0 = - \frac{2(-1)^N}{\pi \eta L_F Z(z)}\, e
^{- \langle \hat{\Phi}^2_{-1} (u_0(z))\rangle _0}
 \cos \left ( 2k_F \tilde{z}(z) - \sqrt{2}b_1 (u_0(z)) \right) e^{-\langle \hat{\Phi} ^2 _1
(u_0(z)) \rangle _0}.
\end{eqnarray}

Note that no Friedel oscillations exist in the composition part because
$\langle \delta \hat{\sigma} (z) \rangle \equiv 0$.

With the renormalized value of the coupling constant $K_{-1} \rightarrow K^*_{-1}$,
the Friedel oscillations are

\begin{eqnarray}\label{48}
\fl \langle \delta \hat{\rho}_F(z) \rangle _0 =
 - \frac{(-1)^N}{\pi L_F} \cos \left [ 2 k_F \tilde{z}(z)- \sqrt{2} b_1 (u_0 (z)) \right]
 \left( \frac{r}{2}\right) ^{ \frac{K_1 + K^*_{-1}}{2}-1}\, Z(z)^{-\left(\frac{K_1+ K_{-1}
^*}{2}+1\right)}
\end{eqnarray}
giving a boundary exponent

\begin{eqnarray}\label{49}
\nu= \frac{K_1+ K^*_{-1}}{4} + \frac{1}{2}.
\end{eqnarray}
Again, this is different from the corresponding exponent

\begin{eqnarray}\label{50}
\nu= \frac{K_\rho+ K^*_{\sigma}}{2}
\end{eqnarray}
of the BLL \cite{FG95,WVP96,VYG00}.

The result equation (\ref{49}) also applies to spinfull one-dimensional fermions ($\nu =1
\rightarrow
c$: charge degrees of freedom, $\nu =-1 \rightarrow s$: spin degrees of freedom) and is thus
applicable to harmonically trapped one-dimensional electrons.

\section{Summary}
We have calculated the quantum interference phenomenon of Friedel oscillations for the 
Tomonaga-Luttinger model of interacting
one-dimensional fermionic atoms trapped in a harmonic potential. We used 
bosonization techniques
and an asymptotic representation of the density operator in terms of a particular phase
field. The result shows that the boundary exponent for the decay of Friedel oscillations
away from the classical boundaries differs from the result for bounded Luttinger liquids
where the fermions are confined between hard walls.

\ack We gratefully acknowledge helpful discussions with F.
Gleisberg and W. P. Schleich and financial help by Deutsche
Forschungsgemeinschaft.

\end{document}